\documentclass[]{spie}  

\usepackage{amsmath,amsfonts,amssymb}
\usepackage{graphicx,verbatim}
\usepackage{booktabs}
\usepackage{multirow}
\usepackage[colorlinks=true, allcolors=blue]{hyperref}

\usepackage{etoolbox}

\AtBeginEnvironment{thebibliography}{%
  \renewcommand{\textbf}[1]{#1}%
  \let\bfseries\mdseries%
  \let\bf\mdseries%
}

\title{Dual-Phase Cross-Modal Contrastive Learning for CMR-Guided ECG Representations for Cardiovascular Disease Assessment}

\author[a,b,c]{Laura Alvarez-Florez$^*$}
\author[c]{Angel Bujalance-Gomez$^*$}
\author[c]{Femke Raijmakers}
\author[g]{Samuel Ruiperez-Campillo}
\author[c]{Maarten Z. H. Kolk}
\author[c]{Jesse Wiers}
\author[g]{Julia Vogt}
\author[f]{Erik J. Bekkers}
\author[a,b,e,h]{Ivana Išgum}
\author[c]{Fleur V. Y. Tjong}

\affil[a]{Department of Biomedical Engineering and Physics, Amsterdam University Medical Center, The Netherlands}
\affil[b]{Quantitative Healthcare Analysis group, University of Amsterdam, The Netherlands}
\affil[c]{Department of Clinical and Experimental Cardiology, Amsterdam University Medical Center, The Netherlands}
\affil[e]{Department of Radiology and Nuclear Medicine, Amsterdam University Medical Center, The Netherlands}
\affil[f]{Amsterdam Machine Learning Lab, University of Amsterdam, The Netherlands}
\affil[g]{Department of Computer Science, ETH Zurich, Switzerland}
\affil[h]{Department of Radiology, Mayo Clinic, Rochester, United States of America}

\authorinfo{Corresponding author: Laura Alvarez-Florez (l.alvarezflorez@amsterdamumc.nl)}

\begin{document} 
\maketitle

\begin{center}
$^*$ These authors contributed equally to this work.
\end{center}

\begin{abstract}
Cardiac magnetic resonance imaging (CMR) offers detailed evaluation of cardiac structure and function, but its limited accessibility restricts use to selected patient populations. 
In contrast, the electrocardiogram (ECG) is ubiquitous and inexpensive, and provides rich information on cardiac electrical activity and rhythm, yet offers limited insight into underlying cardiac structure and mechanical function. 
To address this, we introduce a contrastive learning framework that improves the extraction of clinically relevant cardiac phenotypes from ECG by learning from paired ECG-CMR data. Our approach aligns ECG representations with 3D CMR volumes at end-diastole (ED) and end-systole (ES), with a dual-phase contrastive loss to anchor each ECG jointly with both cardiac phases in a shared latent space. Unlike prior methods limited to 2D CMR representations with or without a temporal component, our framework models 3D anatomy at both ED and ES phases as distinct latent representations, enabling flexible disentanglement of structural and functional cardiac properties. Using over 34,000 ECG-CMR pairs from the UK Biobank, we demonstrate improved extraction of image-derived phenotypes from ECG, particularly for functional parameters ($\uparrow$ 9.2\%), while improvements in clinical outcome prediction remained modest ($\uparrow$ 0.7\%). This strategy could enable scalable and cost-effective extraction of image-derived traits from ECG. The code for this research is publicly available.\footnote{\url{https://github.com/qurAI-amsterdam/dual-phase-contrastive-learning}}

\end{abstract}


\section{INTRODUCTION}
\label{sec:intro}  

Cardiovascular diseases remain the leading global cause of mortality, accounting for a substantial proportion of preventable deaths worldwide \cite{di2024heart}. Many of these conditions, including heart failure, and ischemic heart disease, develop gradually and may remain clinically silent for extended periods, with structural and functional deterioration occurring sometimes before the onset of overt symptoms \cite{sabbah2017silent,cohn2003silent}. This underscores the need for tools that enable precise characterization of cardiac pathology at early stages, where timely intervention may alter disease trajectories and improve outcomes.

Advanced imaging modalities such as cardiac magnetic resonance imaging (CMR) provide comprehensive assessment of cardiac anatomy, function, and tissue properties, and are widely regarded as reference standards for diagnosing a broad range of cardiovascular diseases. However, the cost, complexity, and infrastructure requirements associated with CMR limit its availability to specialized centers, restricting its use to selected patient populations and reducing its feasibility for large scale screening or longitudinal monitoring. In contrast, the electrocardiogram (ECG) is non invasive, inexpensive, and ubiquitously available across a wide range of healthcare settings, from tertiary hospitals to primary care and community based clinics. ECGs are routinely used for the detection of electrical abnormalities such as arrhythmias and conduction disturbances, and form the backbone of cardiovascular screening in many populations. Nevertheless, their ability to capture detailed information about cardiac structure and mechanical function is limited \cite{yong2025non}. Bridging this gap between the accessibility of ECG and the rich functional and structural information provided by CMR brings an opportunity for scalable and cost effective cardiovascular assessment.

Recent advances in deep learning have renewed interest in leveraging ECG data for more comprehensive cardiovascular assessment \cite{liu2021deep, poterucha2025detecting,wang2025tac}. 
While most ECG-based approaches rely on ECG signals alone, multimodal representation learning offers a way to incorporate complementary information from other data sources by embedding heterogeneous modalities into a shared latent space. 
In this setting, contrastive learning has emerged as a common strategy for multimodal alignment, encouraging embeddings from corresponding multimodal pairs to be close in the latent space while separating non-corresponding pairs.
In particular, when applied to ECG and cardiac imaging, this formulation enables ECG representations to be enriched using image-derived information during training, facilitating the learning of embeddings that reflect underlying cardiac structure and function. 
As a consequence, such approaches have the potential to provide a scalable and non-invasive strategy for enabling the extraction of structural and functional cardiac information from ECG alone at inference time.

Previous works have shown that contrastive alignment between ECG and CMR imaging can improve the recovery of structural features from ECG, including ventricular volumes, myocardial mass, and global cardiac geometry and function \cite{Radhakrishnan2023, DingZhengyaoandHu2024, Turgut2025}. 
However, the majority of existing methods rely on simplified image representations, typically using a single two dimensional slice or two dimensional temporal sequences. 
While these representations capture some aspects of cardiac appearance or motion, they may not fully encode the three dimensional geometry of the heart or the spatial relationships between cardiac structures. 
As a result, important aspects of cardiac structure and function may be lost during contrastive learning alignment, particularly those related to global shape, regional remodeling, and volumetric changes across the cardiac cycle. 
This limitation is especially relevant for functional phenotypes, where subtle differences between end diastolic (ED) and end systolic (ES) states reflect myocardial contractility and deformation. 
Capturing such information requires representations that preserve three dimensional anatomy and explicitly model phase dependent cardiac dynamics.

In this work, we extend over prior efforts \cite{DingZhengyaoandHu2024, Turgut2025} by explicitly aligning ECG representations with comprehensive 3D CMR volumes at both ED and ES phases using a dual-phase contrastive learning approach. This design encourages the ECG encoder to capture physiologically meaningful features by jointly modeling structural and functional cardiac dynamics across both imaging phases. We assess its effectiveness for image-derived phenotypes and clinical outcomes.

\section{METHODS}

\subsection{Data}

For training and development, we used data from the UK Biobank \cite{petersen2016uk}, selecting 63,448 participants with ECG records, from which 34,142 had paired CMR acquisitions obtained on the same day. Each ECG consisted of a 10-second, 12-lead recording sampled at 500 Hz. The signals were preprocessed using a standard pipeline including high-pass filtering at 0.5 Hz to remove baseline wander, notch filtering at 60 Hz to suppress power-line interference, Savitzky-Golay filtering for baseline correction and smoothing, and lead-wise normalization to account for amplitude variability across leads.

The corresponding CMR data comprised short-axis cine stacks with up to 12 slices and 50 temporal frames covering a full cardiac cycle. \cite{petersen2016uk}
For each subject, six central slices were selected to focus on the left ventricular region while reducing variability associated with basal and apical slices. 
For each subject, six central slices were selected, which encompass the left ventricular myocardium across subjects, while excluding the more basal and apical regions that are more prone to through plane motion and anatomical variability.
Two cardiac phases, ED and ES, were extracted from each cine sequence. To standardize spatial inputs and reduce background variability, CMR volumes were cropped around the heart using bounding boxes derived from an automated cardiac segmentation model \cite{simonyan2015deepconvolutionalnetworkslargescale}.
Bounding boxes were dilated in each spatial dimension to ensure full myocardial coverage and subsequently zero-padded to obtain consistent input dimensions across subjects. 


The CMR derived phenotypes and cardiovascular outcome labels provided by the UK Biobank were used as reference standards for all prediction tasks in this study. The CMR phenotypes captured key measurements of cardiac structure and function, including ventricular volumes, myocardial mass, myocardial strain, and ejection fraction. Cardiovascular outcomes comprised coronary artery disease, atrial fibrillation, sudden cardiac death, myocardial infarction, heart failure, and cardiomyopathy.

\begin{figure}
    \centering
       \includegraphics[width=\linewidth]{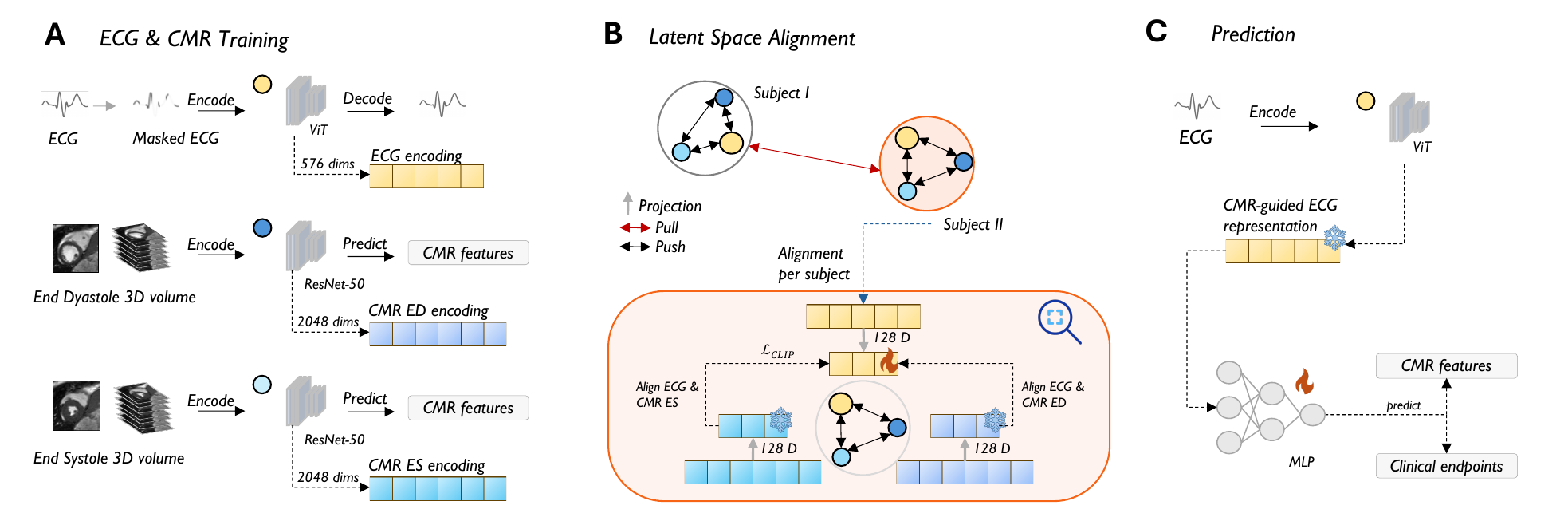}
\caption{Overview of the method. (A) ECG and CMR training: A ViT is trained on ECGs using masked autoencoding, while separate CNN encoders for ED and ES CMR volumes are trained to predict imaging phenotypes. (B) Latent space alignment: ECG and both CMR ES and ED embeddings are aligned simultaneously via a contrastive loss, enabling the ECG encoder to capture structural and temporal cardiac features. (C) Downstream prediction: After contrastive learning, CMR-guided ECG representations are input for an MLP trained to predict CMR phenotypes and clinical endpoints.}
\label{fig:method-main}
\end{figure}

\subsection{Method}



Our deep learning framework performs cardiac phenotyping and predicts clinical outcome leveraging paired ECG–CMR data during training while enabling ECG-only inference. 
The framework consists of three stages: modality-specific encoder pretraining, dual-phase cross-modal contrastive alignment, and downstream phenotype and outcome prediction. 
ECG signals are encoded using a Vision Transformer (ViT) trained in a self-supervised manner, while two separate 3D convolutional encoders are trained on end-diastolic and end-systolic CMR volumes with supervised phenotype prediction. 
The resulting representations are aligned using a dual-phase contrastive learning strategy and the learned ECG embeddings are subsequently used for prediction of CMR-derived phenotypes and clinical outcomes. 
An overview of the proposed approach is shown in Figure~\ref{fig:method-main}.

\textbf{ECG Encoder Training}
The ECG ViT encoder  \cite{dosovitskiy2020image} is trained in a self-supervised manner using a random masking and reconstruction objective. 
During training, segments of the 12-lead ECG signal are randomly masked, and the model is optimized to reconstruct the original signal by minimizing the mean squared error (MSE) between the reconstructed and original waveforms. 
This objective encourages the encoder to learn robust representations of ECG morphology and structure. 
The architecture follows the implementation proposed by Ding et al. \cite{DingZhengyaoandHu2024}, with an embedding dimension of 576, 8 attention heads, 12 transformer layers, and approximately 24 million parameters.

\textbf{CMR Encoder Training}
CMR volumes are encoded using a 3D ResNet-50 architecture \cite{he2016deep}, trained in a supervised manner to regress clinically relevant CMR-derived phenotypes from the UK Biobank. 
In this implementation, two separate CMR encoders are trained, one operating on ED volumes and the other on ES volumes, allowing phase-specific representations to be learned independently. 
Each encoder outputs a 2048-dimensional latent embedding. 
The supervised objective provides explicit guidance for learning representations that capture cardiac anatomy and function. 
The predicted phenotypes include volumetric, mass, and functional measures such as ventricular volumes, myocardial mass, and strain. 
Training is performed using MSE loss and the AdamW optimizer \cite{loshchilov2019decoupledweightdecayregularizationAdamW}, with a cosine annealing learning rate schedule that includes a linear warm-up phase over the first 40 epochs followed by a half-cycle cosine decay. 
Models are trained for up to 200 epochs, with early stopping applied after 15 epochs without validation improvement.

\textbf{Contrastive Cross-Modal Alignment}
Following modality-specific pretraining, ECG and CMR embeddings are aligned using a cross-modal contrastive learning objective inspired by CLIP-style representation learning \cite{radford2021learningtransferablevisualmodels}. A two-layer projection head with a hidden dimension of 512 maps embeddings from each modality into a shared 128-dimensional latent space.

During contrastive training, the CMR encoders are kept frozen to preserve their phenotype-informed representations, while the ECG encoder and projection heads are optimized. Positive pairs consist of ECG and CMR embeddings from the same subject, whereas negative pairs are sampled from different subjects  \cite{DingZhengyaoandHu2024}. The proposed dual-phase formulation encourages the ECG encoder to simultaneously align with both ED and ES CMR representations, thereby embedding complementary structural and functional cardiac information. The contrastive loss is defined as:

\begin{equation}
\mathcal{L}_{a\to b}
=
-\,\frac{1}{N}
\sum_{i=1}^{N}
\log
\frac{
  \exp\bigl(\operatorname{sim}(z_i^a,\,z_i^b)/\tau\bigr)
}{
  \sum_{k=1}^{N}\exp\bigl(\operatorname{sim}(z_i^a,\,z_k^b)/\tau\bigr)
}
\end{equation}

where \(z^a\) and \(z^b\) are the embeddings from any modality pair \((a,b)\), and \(\tau\) is a temperature parameter. The losses are combined using a weighted sum:

\begin{equation}
\mathcal{L}_{\text{dual-phase}}
=
\tfrac{1}{3}\,\bigl(
  \mathcal{L}_{\text{ECG}\to \mathrm{CMR}_{\mathrm{ED}}}
+ \mathcal{L}_{\text{ECG}\to \mathrm{CMR}_{\mathrm{ES}}}
+ \mathcal{L}_{\mathrm{CMR}_{\mathrm{ED}}\to \mathrm{CMR}_{\mathrm{ES}}}
\bigr)
\end{equation}

This formulation extends the CLIP-style \cite{radford2021learningtransferablevisualmodels} contrastive loss to simultaneously align ECG representations with both ED and ES CMR volumes.

\textbf{Prediction of CMR Phenotypes and Clinical Outcomes} After cross-modal alignment, the enriched ECG latent representations trained with contrastive learning are used as input to an MLP for downstream prediction of CMR-derived phenotypes and clinical outcomes. Separate MLPs are trained for each phenotype and clinical outcome to allow task-specific optimization while sharing a common ECG encoder. 
The predicted CMR phenotypes included key structural and functional measures: volumes, myocardial mass, wall thickness, ejection fraction, cardiac output, stroke volume, and global circumferential, longitudinal and radial strain, for the right and left ventricles.
Each MLP consists of three fully connected layers with GELU \cite{hendrycks2016gaussian} activations. 
In addition to imaging derived phenotypes, we evaluate the ability of the ECG representations to predict major cardiovascular disease outcomes. Specifically, we consider coronary artery disease, atrial fibrillation, sudden cardiac death,  myocardial infarction, heart failure, and cardiomyopathy.


\begin{table}[t]
\centering
\resizebox{\textwidth}{!}{%
\begin{tabular}{lcccccc}
\toprule
\multicolumn{7}{c}{\textbf{Volumetric and Mass Phenotypes ($R^{2} \uparrow$)}} \\
\midrule
\textbf{Contrastive Learning Model} & LVDV & LVSV & LVMM & RVDV & RVSV & LVMT \\
\midrule
ECG (baseline) &
$0.606 \pm 0.033$ & $0.548 \pm 0.020$ & $0.640 \pm 0.006$ &
$0.610 \pm 0.008$ & $0.549 \pm 0.029$ & $0.626 \pm 0.014$ \\

ECG 2D CL &
$\mathbf{0.667 \pm 0.006}$ &
$\mathbf{0.639 \pm 0.011}$ &
$\underline{0.681 \pm 0.011}$ &
$\mathbf{0.664 \pm 0.008}$ &
$0.607 \pm 0.009$ &
$\underline{0.685 \pm 0.009}$ \\

ECG 2D+time CL &
$0.629 \pm 0.015$ & $0.579 \pm 0.009$ & $0.673 \pm 0.006$ &
$0.637 \pm 0.016$ & $0.598 \pm 0.008$ & $0.654 \pm 0.005$ \\

ECG 3D-ED CL &
$\underline{0.655 \pm 0.008}$ &
$0.612 \pm 0.005$ &
$0.677 \pm 0.005$ &
$\underline{0.642 \pm 0.018}$ &
$\underline{0.616 \pm 0.005}$ &
$\mathbf{0.688 \pm 0.014}$ \\

ECG 3D CL dual-phase - proposed &
$0.649 \pm 0.025$ &
$\underline{0.624 \pm 0.006}$ &
$\mathbf{0.684 \pm 0.018}$ &
$0.648 \pm 0.011$ &
$\mathbf{0.632 \pm 0.011}$ &
$0.673 \pm 0.016$ \\

\midrule
\multicolumn{7}{c}{\textbf{Functional and Strain Phenotypes ($R^{2} \uparrow$)}} \\
\midrule
\textbf{Contrastive Learning Model} & LVGCS & LVGLS & LVGRS & LVEF & LVCO & RVEF \\
\midrule

ECG (baseline) &
$0.525 \pm 0.011$ & $0.464 \pm 0.020$ & $0.478 \pm 0.027$ &
$0.482 \pm 0.014$ & $0.550 \pm 0.009$ & $0.512 \pm 0.006$ \\

ECG 2D CL &
$\underline{0.603 \pm 0.022}$ &
$0.559 \pm 0.022$ &
$0.568 \pm 0.012$ &
$0.531 \pm 0.007$ &
$0.595 \pm 0.027$ &
$0.534 \pm 0.013$ \\

ECG 2D+time CL &
$0.587 \pm 0.011$ &
$0.549 \pm 0.008$ &
$0.554 \pm 0.019$ &
$0.520 \pm 0.033$ &
$0.595 \pm 0.032$ &
$\underline{0.566 \pm 0.017}$ \\

ECG 3D-ED CL &
$0.568 \pm 0.015$ &
$\underline{0.560 \pm 0.028}$ &
$0.575 \pm 0.019$ &
$\underline{0.536 \pm 0.038}$ &
$\mathbf{0.647 \pm 0.009}$ &
$0.500 \pm 0.083$ \\

ECG 3D CL dual-phase - proposed &
$\mathbf{0.619 \pm 0.014}$ &
$\mathbf{0.567 \pm 0.007}$ &
$\mathbf{0.582 \pm 0.008}$ &
$\mathbf{0.572 \pm 0.004}$ &
$\underline{0.637 \pm 0.012}$ &
$\mathbf{0.579 \pm 0.016}$ \\

\midrule
\multicolumn{7}{c}{\textbf{Clinical Endpoints (AUROC $\uparrow$)}} \\
\midrule
\textbf{Contrastive Learning Model} & CAD & AF & SCD & HF & MI & CMP \\
\midrule

ECG (baseline) &
$68.552 \pm 0.001$ & $73.800 \pm 0.001$ & $62.818 \pm 0.002$ &
$76.836 \pm 0.002$ & $68.862 \pm 0.001$ & $82.667 \pm 0.013$ \\

ECG 2D CL &
$\underline{69.853 \pm 0.074}$ &
$\mathbf{74.406 \pm 0.312}$ &
$62.793 \pm 0.345$ &
$\underline{79.250 \pm 0.193}$ &
$\mathbf{70.379 \pm 0.189}$ &
$83.333 \pm 0.325$ \\

ECG 2D+time CL &
$69.228 \pm 0.290$ &
$73.849 \pm 0.132$ &
$61.867 \pm 0.236$ &
$77.401 \pm 0.212$ &
$69.635 \pm 0.077$ &
$83.889 \pm 0.806$ \\

ECG 3D-ED CL &
$\mathbf{69.872 \pm 0.238}$ &
$\underline{74.304 \pm 0.129}$ &
$62.848 \pm 0.467$ &
$\mathbf{79.676 \pm 0.221}$ &
$\underline{70.348 \pm 0.382}$ &
$\mathbf{85.593 \pm 0.052}$ \\

ECG 3D CL dual-phase - proposed &
$69.748 \pm 0.083$ &
$73.888 \pm 0.100$ &
$\mathbf{63.300 \pm 0.263}$ &
$77.441 \pm 0.226$ &
$69.955 \pm 0.319$ &
$83.111 \pm 0.327$ \\

\bottomrule
\end{tabular}%
}
\caption{
Comparison of ECG based models for cardiac phenotype and clinical outcome prediction, trained under different representation learning settings (rows). All models use ECG only at inference, with imaging information leveraged exclusively during training. The baseline ECG model refers to an ECG only model trained without cross modal contrastive alignment to CMR data. Best value in \textbf{bold}; second best is \underline{underlined}. 
Abbreviations: CL = contrastive learning, LVDV = LV ED volume, LVSV = LV stroke volume, LVMM = LV myocardial mass, RVDV = RV ED volume, RVSV = RV stroke volume, LVMT = LV mean myocardial thickness, LVGCS = LV global circumferential strain, LVGLS = LV global longitudinal strain, LVGRS = LV global radial strain, LVEF = LV ejection fraction, LVCO = LV cardiac output, RVEF = RV ejection fraction, CAD = coronary artery disease, AF = atrial fibrillation, SCD = sudden cardiac death, HF = heart failure, MI = myocardial infarction, CMP = cardiomyopathy.
}
\label{tab:full_stacked}
\end{table}

\section{EXPERIMENTS AND RESULTS}

\begin{figure}[h]
    \centering
    \includegraphics[width=1\linewidth]{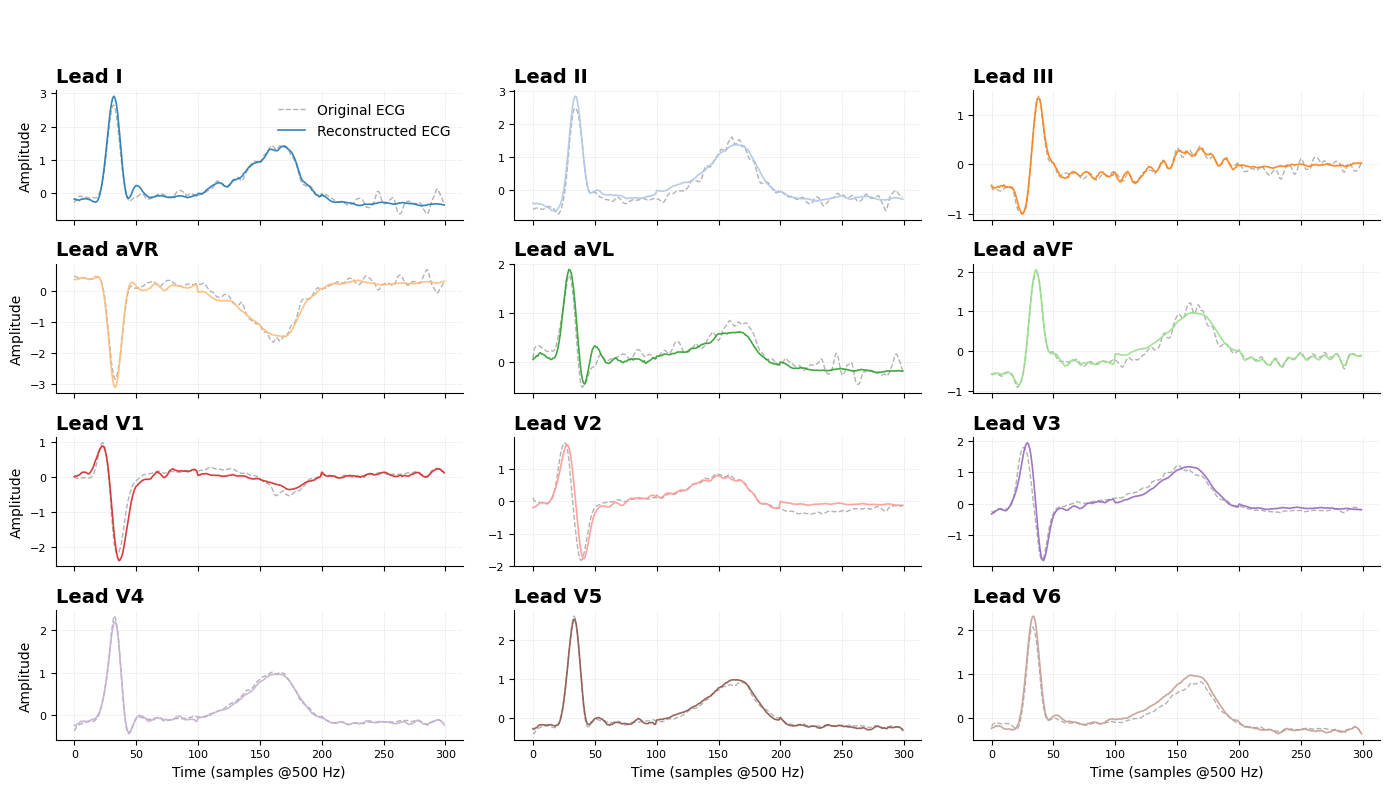}
    \caption{Reconstructed ECG signals for one subject are shown alongside the original 12 lead recordings. The reconstructions display smoother temporal patterns and reduced noise while preserving key waveform characteristics.}
    \label{fig:ECG_reconstruction_576_embeddings_dims}
\end{figure}



To evaluate our dual phase contrastive learning framework for predicting CMR derived cardiac phenotypes and clinical outcomes from ECG signals, we perform multiple experiments investigating how CMR dimensionality, phase composition, and ECG encoder capacity influence performance across tasks.

First, we compare multiple CMR representation strategies within the contrastive learning framework to assess whether increasing spatial dimensionality and incorporating phase-specific volumetric information lead to more informative ECG representations and consistent performance across tasks. The evaluated models include: a baseline ECG model without cross-modal contrastive learning, a 2D ResNet-50 encoding a single mid-ventricular slice at ED, a 2D ResNet-50 combined with an LSTM to encode a single mid-ventricular slice across the full cardiac cycle, a 3D ResNet-50 processing a full ED volume, and the proposed dual-phase approach employing two 3D ResNet-50 encoders at end diastole and end systole. 
Performance was evaluated on twelve CMR phenotypes and six clinical endpoints (see Table~\ref{tab:full_stacked}). 

For all experiments, the following data splits were used. For ECG reconstruction, 47,527 samples were used for training, 10,185 for validation, and 10,185 for testing, with data divided between the sets on the subject level. For subjects with paired ECG–CMR data, the datasets used for contrastive learning and CMR-guided prediction were split into 23,877 patients for training, 5,121 for validation, and 5,144 for testing. For downstream prediction, separate MLP were trained for each task, with training sets for clinical outcome prediction balanced to include equal numbers of positive and negative cases.

The results showed that, for the prediction of volumetric phenotypes, no model demonstrated consistently superior performance across metrics. 
The 2D model performed best on deriving cardiac volumes, while 3D-ED outperformed others on mean wall thickness and cardiac output. 
For the prediction of functional phenotypes, the proposed dual-phase 3D contrastive learning model matched or exceeded other approaches (average $\uparrow$ 9.2\% over baseline ECG), particularly for ejection fraction and all strain measurements. For clinical outcomes, differences across models were modest. 
Although 3D-based models achieved the highest AUROC for several endpoints, improvements were less pronounced (average $\uparrow$ 0.7\% over baseline ECG) than for imaging phenotypes, suggesting limitations in clinical risk prediction from CMR-derived structure alone.

To further investigate the role of ECG encoder capacity, we evaluated four Vision Transformer variants  \textit{Tiny, Small, Medium} and \textit{Base} trained using the same masked reconstruction objective on raw 12-lead ECG signals.
Performance was evaluated on general demographic targets, including age and sex, as well as ECG-derived information, specifically the number and duration of QRS complexes. 
The impact of ECG encoder capacity on representation quality is reflected in the results summarized in Table~\ref{tab:ECG_ablation_perf}, and model configurations and parameter counts in Table~\ref{tab:ECG_ablation_models}. Among the evaluated models, the Medium size model achieved the best QRS-duration $R^{2}$ (0.67) and the highest sex AUC (86.7\%) while requiring fewer parameters than the Base model. 
Figure~\ref{fig:ECG_reconstruction_576_embeddings_dims} exemplifies the reconstructions obtained for one subject form the dataset with the Medium VIT model.

All experiments presented in this evaluation were conducted on NVIDIA H100 GPUs.

\begin{table}[h!]
        \centering
    \begin{tabular}{lcccc}
    
        \toprule
        \multirow{2}{*}{VIT Model} & Sex & Age & Number of QRS in ECG & QRS duration\\
                               & AUC [\%] & $R^{2}$ & $R^{2}$ & $R^{2}$\\
        \midrule
        Tiny   & 82.35 & 0.16 & 0.25 & 0.53\\
        Small  & 85.54 & 0.21 & 0.40 & 0.61\\
        Medium & \textbf{86.74} & \underline{0.44} & \underline{0.44} & \textbf{0.67}\\
        Base   & \underline{85.80} & \textbf{0.47} & \textbf{0.47} & \underline{0.64}\\
        \bottomrule
    \end{tabular}
    \caption{Performance of ViT backbones on general (sex, age) and ECG-derived phenotypes. Best results in \textbf{bold}, second best \underline{underlined}.}
    \label{tab:ECG_ablation_perf}
\end{table}

\begin{table}[h!]
    \centering
    \begin{tabular}{lcccc}
        \toprule
        VIT Model & Layers & Heads & Embedding size & Parameters \\
        \midrule
        Tiny & 12 & 3 & 192 &  5 M\\
        Small & 12 & 6 & 384 &  21 M \\
        Medium & 12 & 8 & 576 &  48 M \\
        Base & 12 & 12 & 768 &  85 M \\
        \bottomrule
    \end{tabular}
    \vspace{0.5em}
    \caption{Network Architecture details of the different ViT models used.}
    \label{tab:ECG_ablation_models}
\end{table}

\section{DISCUSSION}


This study investigated the value of incorporating cardiac structural and functional information from 3D CMR imaging within a dual-phase contrastive learning framework for aligning ECG and CMR representations. By explicitly supervising ECG representations with volumetric cardiac anatomy at both ED and ES phases, our approach aims to enrich ECG embeddings with physiologically meaningful information related to cardiac geometry and function.

Compared to models learning only from 2D CMR with or without temporal features, our contrastive learning approach, which jointly aligns ECG with both 3D volumes at ED and ES CMR phases, achieved the highest or second-highest performance for nearly all CMR phenotypes, particularly for the functional metrics. 
The most pronounced improvements were observed for functional metrics, including ejection fraction and myocardial strain parameters. 
These findings highlight the importance of preserving three-dimensional anatomical context and cardiac dynamics during cross-modal alignment, as simplified image representations may fail to capture global shape changes and volumetric contraction patterns that are critical for functional assessment.
Accordingly, the present work focuses on alignment at ED and ES, which represent the most physiologically informative phases of the cardiac cycle. 
Extending the framework to incorporate additional cine timepoints would require explicit modeling of temporal structure to preserve phase-specific information.
Future work may therefore explore different approaches to leverage temporally resolved CMR data while maintaining the discriminative performance observed with dual phase supervision.

In contrast, improvements in clinical endpoint prediction, including heart failure, coronary artery disease, and myocardial infarction, were more modest. This is likely due to the multifactorial nature of these conditions, which depend not only on cardiac structure and function but also on comorbidities, genetic predisposition, lifestyle factors, and longitudinal disease history. 
While the proposed framework facilitates the extraction of imaging-derived cardiac traits from ECG, further gains in clinical risk prediction may benefit from the integration of additional data sources, such as demographics, laboratory measurements, or clinical records. 
Additionally, a key limitation of this study is that the models are primarily trained on cohorts composed of healthy individuals, with relatively few cases of overt cardiac disease. As a result, generalizability to real world clinical scenarios, particularly among patients with advanced or heterogeneous heart disease, remains largely untested. Future work should therefore prioritize validation in more clinically diverse populations by systematically evaluating the proposed framework on external datasets enriched with heterogeneous cardiac disease.


\section{Conclusions}


In this study, we demonstrated that aligning ECG with full 3D CMR volumes at ED and ES using a dual phase contrastive learning framework improves cross modal representation learning. By incorporating phase specific cardiac structure and function, the proposed approach enables more informative ECG representations for the prediction of CMR derived phenotypes.

\acknowledgments 
This research has been conducted using the UK Biobank Resource under Application Number 24711. This publication is part of the Amsterdam UMC Innovation grant MRI2ECG, and the project DEEP RISK ICD (with project number 452019308) of the research program Rubicon, which is (partly) financed by the Dutch Research Council (NWO), and by a Postdoctoral grant from the Amsterdam Cardiovascular Sciences. The authors gratefully acknowledge funding by the Dutch Heart Foundation (Hartstichting) via the Dekker Grant 2025 for the RESCUE.AI project (Grant No. 03-005-2025-0129).
 







\bibliographystyle{spiebib}
\bibliography{report}

\end{document}